\documentclass[11pt, a4paper]{article}
\usepackage{jheparxiv}
\usepackage[utf8]{inputenc}
\usepackage{amsmath}
\usepackage{amsfonts}
\usepackage{amssymb}
\usepackage{latexsym}
\usepackage{mathrsfs}
\usepackage{braket}		%Bra Ket Notation
\usepackage{graphicx}
\usepackage{color}
\usepackage{xcolor}
\usepackage{slashed}
\usepackage{twistor}
\usepackage[all]{xy}

\newcommand{\sa}{\mathsf{a}}

\renewcommand{\d}{\mathrm{d}}

\subheader{\hfill \texttt{IMPERIAL-TP-TA-2018-03}}

\title{Ambitwistor string vertex operators on curved backgrounds}

\author[a]{Tim Adamo,}
\author[b]{Eduardo Casali}
\author[c]{\& Stefan Nekovar}

\affiliation[a]{Theoretical Physics Group, Blackett Laboratory \\
        Imperial College London, SW7 2AZ, United Kingdom}

\affiliation[b]{Center for Quantum Mathematics and Physics (QMAP) and\\
Department of Physics, University of California, Davis, CA 95616 USA}
        
\affiliation[c]{The Mathematical Institute \\
        University of Oxford, Woodstock Road, OX2 6GG, United Kingdom}

\emailAdd{t.adamo@imperial.ac.uk}
\emailAdd{ecasali@davis.edu}
\emailAdd{nekovar@maths.ox.ac.uk}

\abstract{We present vertex operators for ambitwistor strings around generic Yang-Mills, gravity and NS-NS backgrounds. The requirement that vertex operators lie in the BRST cohomology of the worldsheet theory enforces the appropriate linear equations of motion (as well as gauge fixing conditions) for the respective perturbations in these backgrounds. Due to the nature of ambitwistor strings, no approximation is taken and all calculations around the backgrounds are exact.}

\begin{document}
\notoc

\maketitle\vfill
 
\section{Introduction}

Ambitwistor strings~\cite{Mason:2013sva,Berkovits:2013xba} have many surprising properties; while much attention has rightly been paid to their utility for computing scattering amplitudes, they can also be defined on non-linear background fields~\cite{Adamo:2014wea,Adamo:2018hzd}. On such curved backgrounds the ambitwistor string is described by a chiral worldsheet CFT with free OPEs, which allows for many \emph{exact} computations in these backgrounds, in stark contrast to conventional string theories where an expansion in the inverse string tension is needed (cf., \cite{Fradkin:1985ys,Callan:1985ia,Abouelsaood:1986gd}). For instance, the fully non-linear equations of motion for NS-NS supergravity~\cite{Adamo:2014wea} and gauge theory~\cite{Adamo:2018hzd} emerge as exact worldsheet anomaly cancellation conditions, and ambitwistor strings have been used to compute 3-point functions on gravitational and gauge field plane wave backgrounds~\cite{Adamo:2017sze} correctly reproducing results found with `standard' space-time techniques~\cite{Adamo:2017nia}.

Thus far, only a RNS formalism for the ambitwistor string has been shown to be quantum mechanically consistent at the level of the worldsheet. While pure spinor and Green-Schwarz versions of the ambitwistor string (or deformations thereof) have been defined on curved backgrounds~\cite{Chandia:2015sfa,Chandia:2015xfa,Azevedo:2016zod,Chandia:2016dwr}, it is not clear that they are anomaly-free since only classical worldsheet calculations have been done in these frameworks. In this paper we study the heterotic and type II ambitwistor strings in the RNS formalism, at the expense of only working with NS-NS backgrounds. These backgrounds will be non-linear, and generic apart from constraints imposed by nilpotency of the BRST operator (i.e., anomaly cancellation): the Yang-Mills equations in the heterotic case and the NS-NS supergravity equations in the type II case.

For each of these models, we construct vertex operators in the $(-1,-1)$ picture for all NS-NS perturbations of the backgrounds and investigate the constraints imposed on the operators by BRST closure. In the heterotic model we consider only one such vertex operator whose BRST closure imposes the linearised gluon equations of motion (as well as gauge-fixing conditions) on the perturbation around a Yang-Mills background. In the type II model we consider three vertex operator structures, corresponding to symmetric rank-two tensor, anti-symmetric rank-2 tensor, and scalar perturbations. With a background metric (obeying the vacuum Einstein equations), BRST closure fixes the two tensorial perturbations to be a linearised graviton and $B$-field respectively. On a general NS-NS background (composed of a non-linear metric, $B$-field and dilaton), the three structures are combined into a single vertex operator, whose BRST closure imposes the linearised supergravity equations of motion on the perturbations. 

We comment on the descent procedure for obtaining vertex operators in picture number zero, as well as the prospects for obtaining integrated vertex operators. We also mention some unresolved issues regarding the GSO projection in curved background fields.

\section{Heterotic ambitwistor string}

As a warm up we first describe the vertex operator for a gluon in the heterotic ambitwistor string on a generic Yang-Mills background field since the calculations here are mostly straightforward. This model was defined in a gauge background in~\cite{Adamo:2018hzd}; as usual for ambitwistor strings the worldsheet action is free
\begin{align}\label{wsa2}
S=\frac{1}{2\,\pi}\int_{\Sigma}\Pi_{\mu}\,\dbar X^{\mu}+\frac{1}{2}\,\Psi_{\mu}\,\dbar\Psi^{\mu} +S_{C}\,,
\end{align}
where $\Sigma$ is a closed Riemann surface and $S_{C}$ is the action for a holomorphic current algebra for some gauge group. The bosonic field $X^\mu$ is a worldsheet scalar, and $\Pi_{\mu}$ is its spin $1$ conjugate. The real fermions $\Psi^{\mu}$ are spin $\frac{1}{2}$ fields on the worldsheet. The action \eqref{wsa2} implies free OPEs for the worldsheet fields, along with the usual OPE for a holomorphic worldsheet current algebra:
\begin{align}
\begin{split}\label{OPEs}
&X^{\mu}(z)\,\Pi_{\nu}(w)\sim \frac{\delta^{\mu}_{\nu}}{z-w}\,, \qquad \Psi^{\mu}(z)\,\Psi^{\nu}(w)\sim \frac{\eta^{\mu\nu}}{z-w}\,,\\
&j^{\sa}(z)\,j^{\mathsf{b}}(w)\sim \frac{k\,\delta^{\mathsf{ab}}}{(z-w)^2} + \frac{f^{\mathsf{abc}}\,j^{\mathsf{c}}(w)}{z-w}\,,
\end{split}
\end{align}
where $\eta_{\mu\nu}$ is the $d$-dimensional Minkowski metric, $k$ is the level of the current algebra, and $f^{\mathsf{abc}}$ are the structure constants of the gauge group. At the level of the worldsheet fields dependence on a background gauge field enters through the non-standard gauge transformations of the field $\Pi_{\mu}$. From now on we take the $k\rightarrow 0$ limit to decouple gravitational degrees of freedom from the model~\cite{Berkovits:2004jj,Adamo:2018hzd}.

In addition to the stress-energy tensor $T$, two other (holomorphic) currents are gauged: one is fermionic of spin $\frac{3}{2}$ while the other is bosonic of spin $2$. These currents depend explicitly on the background gauge field $A_{\mu}^{\sa}$; the spin $\frac{3}{2}$ current is
\begin{align}\label{Gcurr}
\mathsf{G}=\Psi^{\mu}\left(\Pi_{\mu}-A^{\mathsf{a}}_{\mu}\,j^{\mathsf{a}}\right)\,,
\end{align}
and the spin $2$ current is
\begin{align}\label{Hcurr}
\mathsf{H} = \Pi^2 - 2\, \Pi^\mu A_{\mu}^{\mathsf{a}} j^\mathsf{a} + A_\mu^\mathsf{a} A^{\mu \mathsf{b}} j^\mathsf{a} j^\mathsf{b} + \Psi^\mu \Psi^\nu F_{\mu\nu}^\mathsf{a} j^\mathsf{a} - \partial\left( \partial_\mu A^{\mu \mathsf{a}} j^\mathsf{a} \right) + f^{\mathsf{a}\mathsf{b}\mathsf{c}} j^\mathsf{c} A^{\mu \mathsf{b}} \partial A_\mu^\mathsf{a}\,.
\end{align}
Here $F_{\mu\nu}^\mathsf{a}$ is the field strength of $A_\mu^{\mathsf{a}}$. It is straightforward to show that these currents obey
\begin{align}\label{Hcurr0}
\mathsf{G}(z)\,\mathsf{G}(w)\sim \frac{\mathsf{H}}{z-w}\,,
\end{align}
without any conditions on the background field.

Constraints on $A_{\mu}^{\sa}$ emerge by requiring the gauging of the currents \eqref{Gcurr} and \eqref{Hcurr} to be quantum mechanically consistent on the worldsheet. Indeed, this gauging leads to the modification of the worldsheet action by ghost systems
\be\label{hghosts}
S\rightarrow S+\frac{1}{2\,\pi}\int_{\Sigma}b\,\dbar c+\tilde{b}\,\dbar\tilde{c}+\beta\,\dbar\gamma\,,
\ee
and an associated BRST charge
\be\label{hBRST}
Q=\oint c\,T +bc\partial c+\gamma\,\mathsf{G}+\frac{\tilde{c}}{2}\,\mathsf{H}+\frac{\tilde{b}}{2}\gamma^2\,,
\ee
for $T$ the full stress-energy tensor (including all ghost and current algebra contributions, except the $(b,c)$ system) and all expressions assumed to be normal-ordered. Here $(b,c)$ are the fermionic ghosts associated to gauging holomorphic worldsheet gravity, $(\beta,\gamma)$ are the bosonic ghosts associated to gauging $\mathsf{G}$, and $(\tilde{b},\tilde{c})$ are the fermionic ghosts associated to gauging $\mathsf{H}$. Both $c,\tilde{c}$ are spin $-1$ while $\gamma$ is spin $-\frac{1}{2}$.

Requiring $Q^2=0$ gives the anomaly cancellation conditions for the theory. The holomorphic conformal anomaly -- controlled entirely through $T$ -- constrains the space-time dimension in terms of the central charge of the current algebra, but puts no restrictions on $A_{\mu}^{\sa}$. However the $\{\mathsf{G},\mathsf{H}\}$ algebra is also anomalous unless it closes: $\mathsf{G}(z)\mathsf{H}(w)\sim 0$. This requirement \emph{does} constrain the background gauge field:
\begin{align}
 \mathsf{G}(z)\mathsf{H}(w)\sim 0 \iff D_{[\mu}F_{\nu\alpha]}^\mathsf{a}=0=D^\mu F_{\mu\nu}^\mathsf{a}\,,
\end{align}
where $D=\partial + A$ is the covariant derivative. These equations are the usual Bianchi identity obeyed by the field strength and the Yang-Mills equations. As expected, vanishing of BRST anomalies imposes on-shell conditions on the background fields.

%%%%%%%%%%%%%%%%%%%%%%%%%%

\subsection{Gluon vertex operator}
\label{glVO}

Our goal is now to describe perturbations of the Yang-Mills background $A_{\mu}^{\sa}$ at the level of vertex operators in the worldsheet CFT. Let $a_\mu^\sa(X)$ be a perturbation of the background. A natural ansatz for an associated vertex operator in the `fixed' picture (i.e., picture number $-1$) is
\begin{align}
V = c\tilde{c}\, \delta(\gamma)\, \Psi^\mu\, a_\mu^\mathsf{a}\, j^\mathsf{a}\,.
\end{align}
This is an admissible vertex operator if it is annihilated by the BRST operator $Q$. Since $V$ is a conformal primary of spin zero, the only interesting contributions to $QV$ come from higher poles in OPEs with the currents \eqref{Gcurr} and \eqref{Hcurr}. Using the free OPEs \eqref{OPEs}, it is straightforward to show that
\begin{align}
\mathsf{G}(z)V(w)\sim - \frac{c\tilde{c}\, \delta(\gamma)\,D^\mu a_\mu^\mathsf{a}\,j^{\sa}(w)}{(z-w)^2}+\cdots\,,
\end{align}
and
\begin{align}
\mathsf{H}(z)V(w)\sim \frac{c\tilde{c}\, \delta(\gamma)\,\Psi^\nu j^{\sa}}{(z-w)^2}\left(D^\mu D_\mu a_\nu^\mathsf{a} + 2  f^{\mathsf{a}\mathsf{b}\mathsf{c}} a^{\mathsf{b} \mu} F^\mathsf{c}_{\mu \nu}\right)(w) + \cdots\,,
\end{align}
where the $+\cdots$ represent single pole terms in the OPE which will not contribute to the action of the BRST charge. 

In particular, these OPEs indicate that
\be\label{gQV}
QV=c\tilde{c}\,\delta(\gamma)\,j^{\sa}\left[\partial\tilde{c}\,\Psi^{\nu}\left(D^{\mu}D_{\mu} a_{\nu}^{\sa}+2 f^{\mathsf{abc}}\,a^{\mathsf{b}\mu}\,F^{\mathsf{c}}_{\mu\nu}\right)-\partial\gamma\,D^{\mu}a_{\mu}^{\sa}\right]\,.
\ee
So requiring $QV=0$ imposes the Lorenz gauge condition ($D^{\mu}a_{\mu}^{\sa}=0$) as well as the linearised Yang-Mills equations
\be\label{linYM}
D^{\mu}D_{\mu} a_{\nu}^{\sa}+2 f^{\mathsf{abc}}\,a^{\mathsf{b}\mu}\,F^{\mathsf{c}}_{\mu\nu}=0\,
\ee
on the perturbation. In other words, the vertex operator lies in the BRST cohomology if and only if $a_{\mu}^{\sa}$ describes an on-shell gluon fluctuation on the non-linear Yang-Mills background.

The standard descent procedure (cf., \cite{Friedan:1985ge,Verlinde:1987sd,Witten:2012bh}) can be used to obtain the gluon vertex operator in zero picture number. To do this, we simply use the standard picture changing operator $\delta(\beta)\mathsf{G}$ to get
\begin{align}
 c\tilde{c}U(w) & =\lim_{z\rightarrow w}\delta(\beta)\mathsf{G}(z)\,V(w) \\
 & =c\tilde{c}\left(\Psi^\mu\Psi^\nu D_\nu a^\mathsf{a}_\mu j^\mathsf{a}+(\Pi^\mu-\mathsf{A}^{\mu\mathsf{a}}j^{\mathsf{a}}) a_\mu^{\mathsf{b}}j^{\mathsf{b}}-f^{\mathsf{abc}}a_{\mu}^{\mathsf{b}}\,j^{\mathsf{c}}\,\partial A^{\mu\mathsf{a}}\right)(w)\,. \label{Dgluon}
\end{align}
An equivalent way to derive $U(w)$ is by linearising the current $\mathsf{H}$ around a Yang-Mills background, keeping in mind that the perturbation $a_{\mu}^{\sa}$ obeys the Lorenz gauge condition.

Further descent into an integrated vertex operator using the $b$-ghost and the stress-energy tensor can be carried out as in the usual string. How to perform the descent using the $\tilde{b}$-ghost and $\mathsf{H}$ current remains an open question, although it is well-known how to do so in a flat background~\cite{Mason:2013sva,Adamo:2013tsa,Ohmori:2015sha}.

%%%%%%%%%%%%%%%%%%%%%%%%%%
%%%%%%%%%%%%%%%%%%%%%%%%%%

\section{Type II ambitwistor string}

We now move on to the type II ambitwistor string on a curved NS-NS background composed of a metric $g_{\mu\nu}$, $B$-field $B_{\mu\nu}$ and dilaton $\Phi$. This model was defined in~\cite{Adamo:2014wea} with worldsheet action
\be\label{IIwsa}
S=\frac{1}{2\,\pi}\int_{\Sigma}\Pi_{\mu}\,\dbar X^{\mu}+\bar\psi_{\mu}\,\dbar\psi^{\mu}+\frac{R_{\Sigma}}{4}\,\mathrm{log}\left(\e^{-2\Phi}\sqrt{g}\right)\,,
\ee
where $(\psi^{\mu},\bar\psi_{\nu})$ is a complex fermion system of spin $\frac{1}{2}$. The final term, proportional to the worldsheet curvature $R_{\Sigma}$, is required to ensure quantum mechanical diffeomorphism invariance, but does not affect local calculations (such as OPEs) since this curvature can always be set to zero in a small neighborhood on the worldsheet. Thus, the OPEs between worldsheet fields remain free and independent of the background fields:
\begin{align}
 \label{OPE}
 X^{\mu}(z)\,\Pi_{\nu}(w)\sim \frac{\delta^{\mu}_{\nu}}{z-w}\,, \qquad \psi^{\mu}(z)\,\bar\psi_{\nu}(w)\sim \frac{\delta^{\mu}_{\nu}}{z-w}\,,
\end{align}
although $\Pi_{\mu}$ does not transform covariantly under a space-time diffeomorphism~\cite{Adamo:2014wea}.

The type II model features the gauging of three additional currents, as well as the holomorphic stress-energy tensor. Two of these are spin $\frac{3}{2}$ fermionic currents,
\begin{align}
\label{GeneralG}
 \mathcal{G}=&\psi^\mu\Pi_\mu + \partial(\psi^\mu\Gamma^\kappa{}_{\mu\kappa})-2\partial(\psi^\mu\partial_\mu\Phi)+\frac{1}{3!}\psi^\mu\psi^\nu\psi^\kappa H_{\mu\nu\kappa}\,,\\
\label{GeneralGbar}
 \bar{\mathcal{G}}=&g^{\mu\nu}\bar\psi_\nu(\Pi_\mu-\Gamma^\kappa{}_{\mu\lambda}\bar\psi_\kappa\psi^\lambda) - g^{\mu\nu}\partial(\bar\psi_\kappa\Gamma^\kappa{}_{\mu\nu})-2\partial(g^{\mu\nu}\bar\psi_\mu\partial_\nu\Phi)+\frac{1}{3!}\bar\psi_\mu\bar\psi_\sigma\bar\psi_\lambda H^{\mu\sigma\lambda}\,,
\end{align}
where $\Gamma^{\kappa}_{\mu\nu}$ are the Christoffel symbols of $g_{\mu\nu}$ and $H_{\mu\nu\sigma}$ is a background three-form. The third current is bosonic of spin $2$, given by\footnote{This expression for $\mathcal{H}$ corrects some typos made in~\cite{Adamo:2014wea}. We have checked that these modifications don't alter any of the results in~\cite{Adamo:2017sze}.}
\begin{align}
\begin{split}
\label{MostGeneralH}
\mathcal{H}=& g^{\mu\nu}\left(\Pi_\mu-\Gamma^\kappa{}_{\mu\lambda}\bar\psi_\kappa\psi^\lambda\right)\left(\Pi_\nu-\Gamma^\kappa{}_{\nu\lambda}\bar\psi_\kappa\psi^\lambda\right) -\frac{1}{2}R^{\kappa\lambda}{}_{\mu\nu}\bar\psi_\kappa\bar\psi_\lambda\psi^\mu\psi^\nu 
 \\ 
& - g^{\mu \nu} \partial\left( \Pi_\rho \Gamma^\rho_{\mu \nu} \right) -\bar\psi_\kappa\partial\psi^\lambda g^{\mu\nu}\partial_\lambda\Gamma^\kappa{}_{\mu\nu} + \psi^\mu \partial_\mu \left( g^{\rho \sigma} \partial (\bar{\psi}_\kappa \Gamma^\kappa_{\rho \sigma} )\right)
\\
&+\frac{1}{2} g^{\mu\nu} H_{\mu \kappa \lambda} \psi^\kappa \psi^\lambda \left(\Pi_\nu - \Gamma^ \rho_{\nu\sigma} \bar{\psi}_\rho \psi^\sigma \right) + \frac{1}{2} \left( \Pi_\mu - \Gamma^\kappa_{\mu\lambda} \bar{\psi}_\kappa \psi^\lambda \right) H_\nu^{\;\; \rho \sigma} \bar{\psi}_\rho \bar{\psi}_\sigma
\\
&+\frac{1}{4} g^{\mu\nu} H_{\mu\kappa\lambda} \psi^\kappa \psi^\lambda H_\nu^{\;\; \rho\sigma} \bar{\psi}_\rho \bar{\psi}_\sigma -\frac{1}{3!} \psi^\mu \bar{\psi}_\nu \bar{\psi}_\kappa \bar{\psi}_\lambda \nabla_\mu H^{\nu\kappa\lambda} - \frac{1}{3!} \bar{\psi}_\mu \psi^\nu \psi^\kappa \psi^\lambda \nabla^\mu H_{\nu\kappa\lambda}
\\
&+ \frac{1}{2} H^{\mu\nu\kappa} \bar{\psi}_\kappa \partial \left( H_{\mu\nu\lambda} \psi^\lambda\right) + \partial \left(H_{\kappa \lambda \nu} \psi^\nu \right) g^{\kappa \sigma} \Gamma^\lambda_{\sigma \rho} \psi^\rho  \,- \partial \left(H_{\kappa \lambda \nu} \psi^\nu  g^{\kappa \sigma} \Gamma^\lambda_{\sigma \rho} \psi^\rho \right)
\\
& - \frac{1}{2} \partial_\sigma H_{\mu\nu\rho} \psi^\nu \psi^\rho \partial g^{\sigma\mu} - \frac{1}{12}  H^{\mu \nu \rho} \partial^2 H_{\mu \nu \rho} + \frac{1}{2} \Gamma^\rho_{\mu\nu} H_{\sigma \lambda \rho} \psi^\sigma \psi^\lambda \partial g^{\mu \nu}
\\
&-2 \partial \left(g^{\mu \nu} \Pi_\mu \partial_\nu \Phi \right) - \partial \left(\bar{\psi}_\kappa \psi^\lambda ( 2 \nabla^\kappa \partial_ \lambda \Phi -2 g^{\mu\nu} \Gamma^\kappa_{\mu \lambda} \partial_\nu \Phi ) \right).
\end{split}
\end{align}
These currents are covariant with respect to target space diffeomorphisms and conformal primaries of the worldsheet CFT. This is despite the fact that they contain various terms which do not appear to be manifestly covariant, due to the requirement of normal-ordering on the worldsheet.

Gauging these currents along with holomorphic worldsheet gravity leads to a BRST operator
\begin{align}\label{IIcQ}
Q=\oint c\,T +bc\partial c+ \frac{\tilde{c}}{2}\,\cH + \bar{\gamma}\,\cG +\gamma\,\bar{\cG}-2\gamma\bar{\gamma}\tilde{b}\,,
\end{align}
where the $(b,c)$, $(\tilde{b},\tilde{c})$, $(\beta,\gamma)$ ghost systems have the same quantum numbers as in the heterotic case, and $(\bar{\beta},\bar{\gamma})$ have the same quantum numbers as their un-barred cousins (i.e., they are bosonic and $\bar{\gamma}$ has spin $-\frac{1}{2}$). The stress tensor can be broken into matter and ghost contributions $T=T_{\mathrm{m}}+T_{\mathrm{gh}}$, with
\begin{align}\label{stress_tensor}
 T_\mathrm{m}= -\Pi_\mu\partial X^\mu -\frac{1}{2}(\psi_\mu\partial\psi^\mu+\psi^\mu\partial\bar\psi_\mu)-\frac{1}{2}\partial^2\log(e^{-2\Phi}\sqrt{g})
\end{align}
for the matter fields and
\begin{align}
 T_{\mathrm{gh}} = \tilde{c}\partial \tilde{b} - 2\tilde{b}\partial \tilde{c} - \frac{3}{2}\beta\partial\gamma - \frac{1}{2}\gamma\partial\beta - \frac{3}{2}\tilde\beta\partial\tilde\gamma - \frac{1}{2}\tilde\gamma\partial\tilde\beta
\end{align}
for the ghost fields, where we again exclude the $(b,c)$ system.

As in the heterotic model, $Q^2=0$ is obstructed by a conformal anomaly and anomalies related to the gauged currents -- in this case $\{\cG,\bar\cG,\cH\}$. The conformal anomaly imposes no constraints on the background fields and is eliminated by selecting the critical space-time dimension $d=10$. The other anomalies vanish if the algebra of currents is quantum mechanically closed:
\be\label{curalg}
\cG(z)\,\cG(w)\sim 0 \sim\bar{\cG}(z)\,\bar{\cG}(w)\,, \qquad \cG(z)\,\bar{\cG}(w)\sim \frac{\cH}{z-w}\,,
\ee
and these conditions impose constraints on the background fields. The requirement that the $\cG(z)\cG(w)$ and $\bar{\cG}(z)\bar{\cG}(w)$ OPEs be non-singular imposes
\be\label{bianchi1}
\partial_{[\mu}H_{\nu\rho\sigma]}=0\,, \quad R_{\mu[\nu\rho\sigma]}=0\,, \quad R_{(\mu\nu)\rho\sigma}=0\,,
\ee
which are the usual Bianchi identities and symmetries of the Riemann tensor of the background metric, along with $\d H=0$. This latter statement indicates that (locally) $H=\d B$; that is, $H$ arises as the field strength of a background $B$-field. 

Dynamical constraints on the background fields emerge from the final closure requirement of \eqref{curalg}, which imposes
\begin{align}
 R+4\nabla_\mu\nabla^\mu\Phi-4\nabla_\mu\Phi\nabla^\mu\Phi-\frac{1}{12}H^2 & =0\,,\nonumber\\
 R_{\mu\nu}+2\nabla_\mu\nabla_\nu\Phi-\frac{1}{4}H_{\mu\rho\sigma}H_\nu{}^{\rho\sigma} & =0\,,\label{SugraEOM}\\
 \nabla_\kappa H^\kappa{}_{\mu\nu}-2H^\kappa{}_{\mu\nu}\nabla_\kappa\Phi & =0\,.\nonumber
\end{align}
These are precisely the field equations for the NS-NS sector of type II supergravity, so vanishing of BRST anomalies enforces the appropriate equations of motion on the background fields.

%%%%%%%%%%%%%%%%%%%%%%%%%%

\subsection{Graviton vertex operator}

To begin, consider the type II model with \emph{only} a background metric $g_{\mu\nu}$ turned on, and let $h_{\mu\nu}(X)$ be a symmetric, traceless perturbation of this metric. A fixed picture vertex operator associated to this perturbation is given by
\begin{align}\label{gravityop}
 V_{h}=c\tilde{c}\,\delta(\gamma)\delta(\bar\gamma)\,\cO_{h}=c\tilde{c}\,\delta(\gamma)\delta(\bar\gamma)\left(\bar{\psi}_\mu\psi^\nu h^\mu{}_\nu-\frac{1}{2}(\partial g_{\mu\nu})h^{\mu\nu}\right).
\end{align}
Note that this contains a quantum correction term proportional to a worldsheet derivative. While this quantum correction vanishes for flat or certain highly symmetric backgrounds (e.g., a plane wave metric written in Brinkmann coordinates~\cite{Adamo:2017sze}), it plays a crucial role on a general background.

For $V_{h}$ to be an admissible vertex operator, it must be annihilated by the BRST operator \eqref{IIcQ}. Since $V_h$ is a conformal primary of spin 0 on the worldsheet, any potential obstructions to its $Q$-closure arise from OPEs between the operator $\cO_h$ and the currents \eqref{GeneralG}, \eqref{GeneralGbar} and \eqref{MostGeneralH} with $H_{\mu\nu\rho}=0=\Phi$. One finds:

\begin{align}
 &\mathcal{G}(z)\,\cO_{h}(w)\sim-\frac{\psi^\nu\,\nabla_{\mu} h^{\mu}{}_{\nu}}{(z-w)^2}(w)+\cdots\,, \\
 &\bar{\mathcal{G}}(z)\,\cO_{h}(w)\sim\frac{g^{\rho\sigma}\bar{\psi}_\mu\,\nabla_\rho h^\mu{}_\sigma}{(z-w)^2}(w)+\cdots\,,
\end{align}
and 
\begin{multline}\label{gback1}
 \frac{\mathcal{H}(z)}{2}\,\cO_h(w)\sim \frac{h^{\mu\nu}R_{\mu\nu}}{(z-w)^3}(w)+\frac{\bar\psi_\alpha\psi^\beta}{2\,(z-w)^2}\left(\nabla_{\kappa}\nabla^{\kappa} h^{\alpha}_{\beta} -  2R^{\alpha}{}_{\sigma \gamma \beta} h^{\sigma \gamma} \right. \\
 \left. -R^{\sigma\alpha} h_{\sigma \beta}  -R^{\sigma}{}_{\beta} h^{\alpha}_{\sigma}+2h^\lambda{}_\beta R_\alpha{}_\lambda\right)(w) +\frac{\partial X^\gamma}{(z-w)^2}\left(\frac{1}{2}h_{\mu\nu}\partial_\gamma R^{\mu\nu}\right. \\
 \left.+\frac{1}{4}\partial_\gamma g^{\mu\nu}(\nabla_{\alpha}\nabla^{\alpha} h_{\mu \nu} - 2 R_{\mu \alpha \beta \nu} h^{\alpha \beta} -R^\lambda_{\; \mu} h_{\lambda \nu}  -R^\lambda_{\; \nu} h_{ \mu \lambda})\right)(w)+\cdots\,,
\end{multline}
where the $+\cdots$ stand for terms which do not contribute to the action of the BRST operator. 

Since the background metric obeys the vacuum Einstein equations ($R_{\mu\nu}=0$), these OPEs imply that
\begin{multline}\label{ggraviton}
QV_{h}=c\tilde{c}\,\delta(\gamma)\delta(\bar\gamma)\bigg[\partial\gamma\,\bar{\psi}_\mu\,\nabla^{\nu} h^\mu{}_\nu-\partial\bar\gamma\, \psi^\nu\,\nabla_{\mu} h^{\mu}{}_{\nu} \\
 \left.+\frac{\partial\tilde{c}\,\bar\psi_\mu\psi^\nu}{2}\left(\nabla_{\alpha}\nabla^{\alpha} h^{\mu}_{\nu} -  2R^{\mu}{}_{\alpha \beta \nu} h^{\alpha \beta}\right) +\frac{\partial\tilde{c}\,\partial g^{\mu\nu}}{4}\,\left(\nabla_{\alpha}\nabla^{\alpha} h_{\mu \nu} - 2 R_{\mu \alpha \beta \nu} h^{\alpha \beta}\right)\right]\,.
\end{multline}
Thus, the OPEs between the vertex operator and the currents $\cG$, $\bar{\cG}$ impose the de Donder gauge condition
\be\label{deDonder}
\nabla^{\mu}h_{\mu\nu}=0\,,
\ee
which is consistent with expectations from the flat background case~\cite{Mason:2013sva}. The OPE between the vertex operator and the current $\cH$ leads to the linearised Einstein equation for a metric perturbation on a vacuum Einstein background:
\be\label{linEin}
\nabla_{\alpha}\nabla^{\alpha} h_{\mu \nu} - 2 R_{\mu \alpha \beta \nu} h^{\alpha \beta}=0\,.
\ee
In other words, requiring $QV_{h}=0$ imposes precisely the physical gauge-fixing and linearised equation of motion for a graviton on the perturbation $h_{\mu\nu}$.

\medskip

What happens when the background $B$-field and dilaton are switched on? Keeping the form \eqref{gravityop} for the vertex operator, it remains to check the action of the \emph{full} (i.e., with $g_{\mu\nu}$, $H_{\mu\nu\rho}$ and $\Phi$) BRST operator \eqref{IIcQ} on $V_h$. The additional background fields do not change the fact that $QV_{h}$ is governed entirely by the OPEs between $\cO_h$ and the currents \eqref{GeneralG}, \eqref{GeneralGbar} and \eqref{MostGeneralH}, although these OPEs are now substantially more complicated. One finds that
\begin{align}
 &\mathcal{G}(z)\,\cO_{h}(w)\sim-\frac{\psi^\nu}{(z-w)^2}\left(\nabla_\mu h^\mu{}_\nu-2h^\mu{}_\nu\partial_\mu\Phi\right)+\cdots \,,\\
 &\bar{\mathcal{G}}(z)\,\cO_{h}(w)\sim\frac{g^{\rho\sigma}\bar{\psi}_\mu}{(z-w)^2}\left(\nabla_\rho h^\mu{}_\sigma-2h^\mu{}_\rho\partial_\sigma\Phi\right)+\cdots\,,
\end{align}
while the OPE between $\cH$ and $\cO_h$ is 
\begin{multline}
 \frac{\cH(z)}{2}\,\cO_{h}(w)\sim \frac{h^{\mu\nu}}{(z-w)^3}\left(R_{\mu\nu}+2\nabla_\mu\nabla_\nu\Phi-\frac{1}{4}H_{\mu\rho\sigma}H_\nu{}^{\rho\sigma}\right) \\
 +\frac{\bar\psi_{\alpha}\psi^{\beta}}{(z-w)^2}\left[h^{\lambda}_{\beta}\left(R^{\alpha}{}_{\lambda}+2\nabla^\alpha\nabla_\lambda\Phi-\frac{1}{4}H^{\alpha}{}_{\rho\sigma}H_\lambda{}^{\rho\sigma}\right) + \frac{1}{2}\left(\nabla_{\lambda}\nabla^{\lambda}h^{\alpha}_{\beta}-2R^{\alpha}{}_{\sigma\rho\beta}h^{\sigma\rho}-R^{\sigma\alpha}h_{\sigma\beta}\right. \right. \\
 -R^{\sigma}{}_{\beta}h^{\alpha}_{\sigma}-h^{\rho}_{\sigma} H_{\beta\rho\kappa}H^{\alpha\sigma\kappa}-2(h^{\alpha}_{\sigma}\nabla_{\beta}\partial^{\sigma}\Phi+h_{\beta\sigma}\nabla^{\alpha}\partial^{\sigma}\Phi+\nabla_{\sigma}h^{\alpha}_{\beta} \partial^{\sigma}\Phi)\Big)\bigg] \\
 +\frac{1}{(z-w)^2}\left[\frac{h_{\mu\nu}}{2}\,\partial\!\left(R^{\mu\nu}+2\nabla^\mu\nabla^\nu\Phi-\frac{1}{4}H^{\mu}{}_{\rho\sigma}H^{\nu\rho\sigma}\right)+\frac{\partial g^{\mu\nu}}{4}\left(\nabla_{\lambda}\nabla^{\lambda}h_{\mu\nu}-2R_{\mu\alpha\beta\nu}h^{\alpha\beta} \right.\right. \\
 -R^{\lambda}{}_{\mu}h_{\lambda\nu}-R^{\lambda}{}_{\nu}h_{\lambda\mu}-h^{\lambda}_{\sigma}H_{\mu\lambda\alpha}H_{\nu}{}^{\sigma\alpha}-2\left(h_{\mu\sigma}\nabla_{\nu}\partial^{\sigma}\Phi-h_{\nu\sigma}\nabla_{\mu}\partial^{\sigma}\Phi+\nabla_{\sigma}h_{\mu\nu}\partial^{\sigma}\Phi\right)\Big)\bigg] \\
 +\frac{\psi^{\rho}\psi^{\sigma}}{2\,(z-w)^2}\left(\nabla_{\nu}h_{\lambda\sigma}\,H_{\rho}{}^{\nu\lambda}+\frac{h^{\alpha\beta}}{2}\,\nabla_{\alpha}H_{\beta\sigma\rho}\right)-\frac{\bar\psi_{\rho}\bar{\psi}_{\sigma}}{2\,(z-w)^2}\left(\nabla_{\nu}h_{\lambda}^{\sigma}\,H^{\rho\nu\lambda}+\frac{h^{\alpha\beta}}{2}\,\nabla_{\alpha}H_{\beta}{}^{\sigma\rho}\right) \\ +\cdots\,,
\end{multline}
where all numerators are evaluated at $w$ on the worldsheet, and $+\cdots$ again denotes terms which will not contribute to the action of the BRST operator.

Using the fact that the background fields obey the non-linear equations of motion \eqref{SugraEOM}, this means that
\begin{multline}\label{NSgraviton}
 QV_{h}=c\tilde{c}\,\delta(\gamma)\delta(\bar\gamma)\bigg[\partial\gamma\,\bar{\psi}_\mu\,(\nabla^{\nu} h^\mu{}_\nu-2h^{\mu}{}_{\nu}\partial^{\nu}\Phi)-\partial\bar\gamma\, \psi^\nu\,(\nabla_{\mu} h^{\mu}{}_{\nu}-2h^{\mu}{}_{\nu}\partial_{\mu}\Phi) \\
 \frac{\partial\tilde{c}}{4}\,\left(2\bar\psi^{\mu}\psi^{\nu}+\partial g^{\mu\nu}\right)\left(\nabla_{\lambda}\nabla^{\lambda}h_{\mu\nu}-2R_{\mu\rho\sigma\nu}h^{\rho\sigma}-R^{\lambda}{}_{\mu}h_{\lambda\nu}-R^{\lambda}{}_{\nu}h_{\lambda\mu}\right. \\
 \left.-h^{\lambda}_{\sigma}H_{\mu\lambda\alpha}H_{\nu}{}^{\sigma\alpha}-2\left(h_{\mu\sigma}\nabla_{\nu}\partial^{\sigma}\Phi+h_{\nu\sigma}\nabla_{\mu}\partial^{\sigma}\Phi+\nabla_{\sigma}h_{\mu\nu}\partial^{\sigma}\Phi\right)\right) \\
 +\frac{\partial\tilde{c}}{2}\left(\psi^{\mu}\psi^{\nu}-\bar{\psi}^{\mu}\bar{\psi}^{\nu}\right)\left(\nabla_{\rho}h_{\lambda\nu}\,H_{\mu}{}^{\rho\lambda}-\frac{h^{\rho\sigma}}{2}\,\nabla_{\rho}H_{\sigma\mu\nu}\right)\bigg]\,,
\end{multline}
where indices are raised and lowered with the background metric. The requirement $QV_{h}=0$ therefore imposes the generalized de Donder gauge condition
\be\label{gdeDonder}
\nabla^{\mu}h_{\mu\nu}=2\,h_{\mu\nu}\partial^{\mu}\Phi\,,
\ee
as well as the linearised equation of motion
\begin{multline}\label{linNSEin}
\nabla_{\lambda}\nabla^{\lambda}h_{\mu\nu}-2R_{\mu\rho\sigma\nu}h^{\rho\sigma}-R^{\lambda}{}_{\mu}h_{\lambda\nu}-R^{\lambda}{}_{\nu}h_{\lambda\mu}-h^{\lambda}_{\sigma}H_{\mu\lambda\alpha}H_{\nu}{}^{\sigma\alpha} \\
-2\left(h_{\mu\sigma}\nabla_{\nu}\partial^{\sigma}\Phi+h_{\nu\sigma}\nabla_{\mu}\partial^{\sigma}\Phi+\nabla_{\sigma}h_{\mu\nu}\partial^{\sigma}\Phi\right)=0\,.
\end{multline}
As desired, this is precisely the linearisation of the symmetric tensor equation from \eqref{SugraEOM} for a metric perturbation.

However, we also obtain an \emph{antisymmetric} constraint from the last line of \eqref{NSgraviton}:
\be\label{skewgrav}
\nabla_{\rho}h_{\lambda\nu}\,H_{\mu}{}^{\rho\lambda}-\frac{h^{\rho\sigma}}{2}\,\nabla_{\rho}H_{\sigma\mu\nu}=0\,.
\ee
From a space-time perspective, this is unexpected: given a symmetric, traceless perturbation $h_{\mu\nu}$, one only expects to obtain the symmetric equation of motion \eqref{linNSEin}. The antisymmetric equation \eqref{skewgrav} arises because the background fields $\{g,H,\Phi\}$ are still treated as fluctuating quantum fields by the worldsheet theory. Indeed, these background fields are functionals of the worldsheet field $X^{\mu}(z)$, which is a full quantum field contributing to all OPEs. 

This means that the perturbation $h_{\mu\nu}$ can backreact on the background geometry, leading to additional constraints. In particular, a metric perturbation sources terms in the antisymmetric equation of motion for the background fields \eqref{SugraEOM}\footnote{The metric perturbation can also source a scalar constraint, but it is easy to see that this vanishes on the support of the background equations of motion.}. At the level of a space-time variational problem, this corresponds to evaluating the space-time action on $\{g+h,H,\Phi\}$ and varying it with respect to all these fields. Projecting the resulting equations of motion onto the parts linear in $h$ gives the symmetric equation \eqref{linNSEin} and the antisymmetric equation \eqref{skewgrav} as well as the trivial scalar constraint.

Consequently, the graviton vertex operator only makes sense in the BRST cohomology in the presence of a background metric. When a full NS-NS background is turned on, $QV_h=0$ leads to the physical gauge-fixing condition \eqref{gdeDonder} and correct equation of motion \eqref{linNSEin}, but also an additional backreaction constraint \eqref{skewgrav}. We will see the resolution of this issue in Section~\ref{NSNSvertex}.

%%%%%%%%%%%%%%%%%%%%%%%%%%

\subsection{B-field vertex operator}

Consider a $B$-field perturbation $b_{\mu\nu}(X)$, which is anti-symmetric ($b_{\mu\nu}=b_{[\mu\nu]}$). As in the graviton case, initially we seek a vertex operator to describe this perturbation on a background metric $g_{\mu\nu}$ alone. Using consistency with the flat space GSO projection as a guide, the candidate vertex operator in the fixed picture is:
\be\label{0bvertex}
V_{b}^{(0)}=\frac{c\tilde{c}}{2}\,\delta(\gamma)\delta(\bar\gamma)\,\left(\psi^{\mu}\psi^{\nu}\,b_{\mu\nu}-\bar{\psi}_{\mu}\bar{\psi}_{\nu}\,b^{\mu\nu}\right)\,.
\ee
It is straightforward to compute the action of the BRST operator $Q$ on $V^{(0)}_{b}$; since the operator is a conformal primary of spin zero with a canonical ghost structure, $QV^{(0)}_b$ is controlled entirely by the OPEs between the terms in brackets in \eqref{0bvertex} and the currents $\cG$, $\bar{\cG}$, $\cH$ (with $H_{\mu\nu\rho}=0=\Phi$).

This leads to
\begin{multline}\label{0bfield}
QV^{(0)}_{b}=c\tilde{c}\,\delta(\gamma)\delta(\bar\gamma)\,\bigg[\partial\gamma\,\bar{\psi}_\nu\,\nabla_{\mu} b^{\mu\nu}+\partial\bar\gamma\, \psi^\nu\,\nabla^{\mu} b_{\mu\nu} \\
 +\frac{\partial\tilde{c}}{4}\left(\psi^{\mu}\psi^{\nu}-\bar{\psi}^{\mu}\bar{\psi}^{\nu}\right)\left(\nabla_{\lambda}\nabla^{\lambda} b_{\mu\nu} -  2R_{\sigma\mu\nu\rho} b^{\sigma\rho}+2R^{\sigma}{}_{\mu} b_{\nu\sigma} \right)\bigg]\,.
\end{multline}
Using the vacuum Einstein equations for the background, $QV^{(0)}_b=0$ imposes the gauge-fixing constraint
\be\label{0bgauge}
\nabla^{\mu}b_{\mu\nu}=0\,,
\ee
as well as the equation of motion
\be\label{0bEOM}
\nabla_{\lambda}\nabla^{\lambda} b_{\mu\nu} -  2R_{\sigma\mu\nu\rho}\,b^{\sigma\rho}=0\,
\ee
on the perturbation. Sure enough, \eqref{0bEOM} is precisely the linearised equation of motion for a $B$-field propagating on a vacuum Einstein background.

\medskip

From our experience with the graviton vertex operator, we know that a $B$-field perturbation in a general NS-NS background will source the linearised scalar and symmetric tensor equations of motion, leading to unwanted constraints on the perturbation. Nevertheless, it is instructive to see how this arises by constructing a vertex operator for the perturbation $b_{\mu\nu}$ with a background metric, $B$-field and dilaton.

It is easy to see that $V^{(0)}_{b}$ is no longer correct in this case; we claim that it must be supplemented by additional terms with non-standard worldsheet ghost structure. To write these terms down, we must bosonize the worldsheet ghost systems $(\beta, \gamma)$ and $(\bar\beta,\bar\gamma)$~\cite{Friedan:1985ge}. Let $\phi$ be a chiral scalar on the worldsheet, and $(\eta,\xi)$ be a pair of fermions of spin $+1$ and $0$, respectively. These fields have OPEs 

\be\label{bgs}
 \phi(z)\,\phi(w)\sim -\ln(z-w)\,, \qquad \eta(z)\,\xi(w)\sim \frac{1}{z-w},
\ee
and are related to the ghosts $(\beta,\gamma)$ by
\be\label{bgs1}
 \gamma=\eta\, e^{\phi}\,, \qquad \beta=e^{-\phi}\,\partial\xi\,,
\ee
using the fact that an exponential of the chiral scalar $\e^{k\phi}$ has spin $-(k+\frac{k^2}{2})$. An additional copy of each system, $\bar\phi$, $(\bar{\eta},\bar{\xi})$ is introduced (with identical statistics) for the $(\bar\beta,\bar\gamma)$ ghost system.

With these bosonized ghost systems, the $B$-field vertex operator on a general NS-NS background is given by
\be\label{bvertex}
V_{b}=V^{(0)}_{b}+\cO_{b}^{(1)}+\bar{\cO}_{b}^{(1)}\,,
\ee
where the additional operators are
\begin{align}
\begin{split}
\mathcal{O}_b^{(1)} & = \frac{c \tilde{c}}{4} \partial \tilde{c} \, \partial \xi \, \e^{-2 \phi} \e^{-\bar{\phi}} \,\psi^\mu H_{\mu\rho\sigma} b^{\rho \sigma}\,,
\\
\bar{\mathcal{O}}_b^{(1)} &= \frac{c \tilde{c}}{4} \partial \tilde{c} \, \partial \bar{\xi} \, \e^{-2 \bar{\phi}} \e^{-{\phi}} \,\bar{\psi}_\mu H^{\mu\rho\sigma} b_{\rho \sigma}\,.
\end{split} 
\end{align}
The fact that these additional operators are required is perhaps not surprising, since the background $B$-field couples to the BRST operator in a manner that is distinctly different to the background metric.

We must now check the action of the BRST operator on $V_b$. While $QV_{b}^{(0)}$ was governed entirely by the OPEs between the currents $\cG$, $\bar{\cG}$ and $\cH$, the same is not true of $QV_b$. This is due to the non-standard ghost structure of $\cO_{b}^{(1)}$, $\bar{\cO}^{(1)}_{b}$. For instance, there are now non-trivial OPEs with the structure constant terms in \eqref{IIcQ} that must be accounted for:
\begin{align}
-2 \tilde{b} \gamma \bar{\gamma}(z)\, \mathcal{O}^{(1)}_b(w) &\sim \frac{c \tilde{c} e^{-\bar{\phi}} \eta }{z-w} \frac{\bar{\psi}_\mu H^{\mu\rho\sigma} b_{\rho \sigma}}{2}+\cdots\,, \label{gauge_cancel1}
\\
-2 \tilde{b} \gamma \bar{\gamma} (z) \bar{\cO}^{(1)}_b(w) &\sim \frac{c \tilde{c} e^{-{\phi}} \bar{\eta} }{z-w} \frac{{\psi}^\mu H_{\mu\rho\sigma} b^{\rho \sigma}}{2}+\cdots\,, \label{gauge_cancel2}
\end{align}
making use of the general rule
%\be\label{bgs2}
%\e^{\pm\phi}(z)\,\e^{k\phi}(w)\sim (z-w)^{\mp k}\,\e^{(k \pm 1)\phi} \pm \frac{(z-w)^{1\mp k}}{(k\pm 1)}\,\partial \e^{(k\pm 1)\phi} +\cdots\,,
%\ee
\begin{align}
 \e^{\pm\phi}(z)\,\e^{k\phi}(w)= (z-w)^{\mp k}:\e^{\pm\phi}(z)\,\e^{k\phi}(w):
\end{align}
for OPEs between exponentials of the chiral scalar. Note that contributions from the expansion of $\e^{\pm\phi}(z)$ are of crucial importance, canceling algebraic contributions to the OPEs
\begin{align}
\bar{\gamma} \mathcal{G}(z)\, V^{(0)}_b(w) &\sim -\frac{c \tilde{c} \e^{-\phi} \bar{\eta}}{z-w} \left( \bar{\psi}_\beta \left(\nabla_\alpha b^{\alpha \beta} - 2b^{\alpha \beta} \partial_\alpha \Phi \right) + \frac{\psi^\mu H_{\mu\rho\sigma} b^{\rho \sigma}}{2}  \right)\,, \label{gauge_cond1}
\\
\gamma \bar{\mathcal{G}}(z)\, V^{(0)}_b(w) &\sim -\frac{c \tilde{c} \e^{-\bar{\phi}} {\eta}}{z-w} \left( {\psi}^\beta \left(\nabla^\alpha b_{\alpha \beta} - 2b_{\alpha \beta} \partial^\alpha \Phi \right) + \frac{\bar{\psi}_\mu H^{\mu\rho\sigma} b_{\rho \sigma}}{2}  \right)\,.\label{gauge_cond2}
\end{align}
Similarly, at every stage of this calculation it is crucial to consider all possible contributions from ghosts to the OPEs. Note that contributions from the stress-energy tensor terms in $Q$ remain trivial, since both $\cO_{b}^{(1)}$ and $\bar{\cO}^{(1)}_{b}$ are conformal primaries of spin zero -- despite their non-trivial ghost structure.

The final result of these calculations is
\begin{multline}\label{bfield}
 QV_b= \frac{c \tilde{c}}{4} \partial \tilde{c}\, \e^{-\phi} \e^{-\bar{\phi}}\bigg[\bar{\psi}_\rho {\psi}^\sigma \left( H^{\mu \alpha \rho} (\d b)_{\mu\alpha\sigma} +  H_{\mu \alpha\sigma} (\d b)^{\mu\alpha\rho}  \right)+ \partial g^{\rho\sigma} \left( H^{\mu \beta}{}_\rho (\d b)_{\mu\beta\sigma} \right) \\
+(\psi^\mu \psi^\nu - \bar{\psi}^\mu \bar{\psi}^\nu)\left(\nabla_{\lambda}\nabla^{\lambda} b_{\mu\nu} -   2R_{\alpha\mu\nu\beta} b^{\alpha\beta} + 2R^\alpha{}_{\mu} b_{\nu \alpha }- 2\partial^\alpha \Phi  \nabla_\alpha b_{\mu\nu} +4 b_{\alpha \mu} \nabla_\nu \partial^\alpha \Phi\right)\bigg] \\
-c \tilde{c}\, \e^{-\phi} \bar{\eta}\,\bar{\psi}_\beta \left(\nabla_\alpha b^{\alpha \beta} - 2b^{\alpha \beta} \partial_\alpha \Phi \right)-c \tilde{c}\, \e^{-\bar{\phi}} \eta\, \psi^\beta \left(\nabla^\alpha b_{\alpha \beta} - 2b_{\alpha \beta} \partial^\alpha \Phi \right) \\
+\frac{c \tilde{c}}{12} \partial \tilde{c}\, \e^{-\phi}\, \partial \e^{-\bar{\phi}}\,H^{\mu\nu\rho} (\d b)_{\mu\nu\rho}- \frac{c \tilde{c}}{12} \partial \tilde{c}\, \partial \e^{-\phi}\, \e^{-\bar{\phi}}\,H^{\mu\nu\rho} (\d b)_{\mu\nu\rho}\,,
\end{multline}
where $(\d b)_{\mu\alpha\sigma}=\partial_\mu b_{\alpha\sigma}+\partial_\alpha b_{\sigma\mu}+\partial_\sigma b_{\mu\alpha}$ and all terms proportional to the background equations of motion \eqref{SugraEOM} have been set to zero. As desired, setting $QV_b=0$ enforces the gauge condition
\be\label{bgauge}
\nabla^\mu b_{\mu\nu} = 2 b_{\mu\nu}\, \partial^\mu \Phi\,,
\ee
along with the linearised equation of motion for a $B$-field perturbation on a NS-NS background:
\be\label{bEOM}
 \nabla_{\lambda}\nabla^{\lambda} b_{\mu\nu} -2 R_{\rho\mu \nu\sigma}\, b^{\rho\sigma}+2R^\sigma{}_{[\mu} b_{\nu] \sigma} -2 \partial^\sigma \Phi\, \nabla_\sigma b_{\mu\nu} + 4 b_{\sigma [\mu}\, \nabla_{\nu]} \partial^\sigma \Phi=0\,.
\ee
We also obtain additional scalar and symmetric backreaction constraints on the perturbation:
\be\label{sbfield}
H_{\mu}{}^{\rho\sigma}\, (\d b)_{\nu\rho\sigma}=0=H\cdot(\d b)\,.
\ee
So as expected, $V_b$ only makes sense in the BRST cohomology on a purely metric background.

\subsection{Dilaton vertex operator}

In usual superstring theory, the form of the dilaton vertex operator~\cite{Kataoka:1990ga} is complicated by the fact that the dilaton couples to the worldsheet action through the Fradkin-Tseytlin term~\cite{Fradkin:1985ys}. A similar mechanism is in play in the ambitwistor string, visible at the level of the BRST charge through the last term in the matter stress-energy tensor \eqref{stress_tensor}. For a scalar perturbation on space-time $\varphi(X)$, the associated ambitwistor string vertex operator is composed of four terms: 
\begin{align}
\label{DilatonVO}
 V_\varphi =  \cO^{(1)}_\varphi + \bar{\cO}^{(1)}_\varphi + \cO^{(2)}_\varphi + \bar{\cO}^{(2)}_\varphi\,,
\end{align}
where
\begin{align}
\cO^{(1)}_\varphi &= - c \tilde{c}\, \partial \tilde{c}\, \partial \xi \,   \e^{-2 \phi}\, \e^{- \bar{\phi}}\,{\psi}^\mu \partial_\mu \varphi\,,
\\
\bar{\cO}^{(1)}_\varphi &= - c \tilde{c}\, \partial \tilde{c}\, \partial \bar{\xi} \, \e^{-2 \bar{\phi}}\, \e^{- {\phi}}\, \bar{\psi}_\mu \partial^\mu \varphi\,,
\\
\cO^{(2)}_\varphi &= 2\, c \tilde{c}\, \partial \e^{-\phi}\, \e^{-\bar{\phi}}\, \varphi\,,
\\
\bar{\cO}^{(2)}_\varphi &=  -2\, c \tilde{c}\, \e^{-\phi}\,  \partial \e^{-\bar{\phi}}\, \varphi\,.
\end{align}
Note that unlike the graviton and $B$-field vertex operators, \eqref{DilatonVO} differs in the flat space limit from other formulae appearing in the literature~\cite{Berkovits:2018jvm}. This is due to our use of a complex fermion system for the spin $\frac{1}{2}$ matter fields on the worldsheet, as opposed to the real fermion system used elsewhere.

Unlike the previous cases, not all constituents of $V_{\varphi}$ are conformal primaries. In particular, the operators $\cO^{(2)}_\varphi$ and $\bar{\cO}^{(2)}_\varphi$ are not primary, so when calculating $QV_{\varphi}$ care must be taken to account for contributions from their OPEs with stress tensor terms in the BRST operator. The relevant OPEs are
\begin{align}
\begin{split}
(c T  + bc\partial c)\!(z)\, \cO^{(2)}_\varphi(w) &\sim - 2\,\frac{c \partial c\, \tilde{c}\, \e^{-\phi } \e^{-\bar{\phi} } }{(z-w)^2}\,\varphi+\cdots\,,
\\
(c T  + bc\partial c)\!(z)\, \bar{\cO}^{(2)}_\varphi(w) &\sim 2\,\frac{c \partial c\, \tilde{c}\, \e^{-\phi } \e^{-\bar{\phi} } }{(z-w)^2}\,\varphi+\cdots\,,
\end{split} 
\end{align} 
so the anomalous conformal weight contributions cancel between the two operators. 

The non-trivial ghost structure of all four contributions in \eqref{DilatonVO} necessitates a careful treatment of the ghost contributions to the action of the BRST operator. On a general NS-NS background, the result is
\begin{multline}\label{dilaton}
 QV_{\varphi}=2c\tilde{c}\,\partial\tilde{c}\left(\partial\e^{-\phi}\,\e^{-\bar{\phi}}-\e^{-\phi}\,\partial\e^{-\bar{\phi}}\right)\left(\nabla_{\mu}\partial^{\mu}\varphi-2\,\partial_{\mu}\Phi\,\partial^{\mu}\varphi\right) \\
 -c\tilde{c}\,\partial\tilde{c}\,\e^{-\phi}\e^{-\bar\phi}\left[\left(\partial g^{\mu\nu}+2\bar{\psi}^{\mu}\psi^{\nu}\right)\,\nabla_{\mu}\partial_{\nu}\varphi+\frac{1}{2}\left(\psi^{\mu}\psi^{\nu}-\bar{\psi}^{\mu}\bar{\psi}^{\nu}\right)\,H_{\mu\nu\sigma}\,\partial^{\sigma}\varphi\right]\,.
\end{multline}
Requiring $QV_{\varphi}=0$ therefore imposes scalar, symmetric and anti-symmetric equations of motion on the perturbation:
\be\label{scalardilaton}
\nabla_{\mu}\partial^{\mu}\varphi-2\,\partial_{\mu}\Phi\,\partial^{\mu}\varphi=0
\ee
\be\label{tensordilaton}
\nabla_{\mu}\partial_{\nu}\varphi=0\,, \qquad  H_{\mu\nu\sigma}\,\partial^{\sigma}\varphi=0\,.
\ee
As expected, only the scalar equation~\eqref{scalardilaton} is the desired one; the two tensor equations~\eqref{tensordilaton} arise from the backreaction of the scalar perturbation on the metric and $B$-field sectors.

However, the situation for the dilaton vertex operator is worse than for the graviton or $B$-field: even with a pure metric background, we still obtain a tensor equation $\nabla_{\mu}\partial_{\nu}\varphi=0$, which over-constrains the perturbation. Although the vertex operator \eqref{DilatonVO} gives the correct scalar equation of motion, its inclusion in the BRST cohomology enforces unphysical constraints on the spectrum

\subsection{NS-NS vertex operator}
\label{NSNSvertex}

For each of the graviton, $B$-field and dilaton vertex operators, we have seen that the associated vertex operator is not in the BRST cohomology of the type II ambitwistor string on a general NS-NS background. While the graviton \eqref{gravityop} and $B$-field \eqref{0bvertex} vertex operators are BRST-closed on the support of the appropriate linearised field equations on a pure gravity background, the dilaton operator is only BRST-closed on the support of additional, unphysical equations for \emph{any} sector of background fields.

These issues are overcome by combining the graviton, $B$-field and dilaton vertex operators into a single NS-NS vertex operator, which simultaneously perturbs each sector of the background. Indeed, from the space-time perspective this is much more natural than exciting a perturbation of one of the fields on its own, since the non-linear equations of motion \eqref{SugraEOM} intertwine all three. This `fat graviton,' sometimes expressed heuristically as $h_{\mu\nu}\oplus b_{\mu\nu}\oplus\varphi$, is the natural perturbation of the NS-NS sector of type II supergravity.

The candidate vertex operator is given by summing together each of three vertex operators constructed above:
\be\label{NSvertex}
V_{\mathrm{NS}}=V_{h}+V_{b}+V_{\varphi}\,,
\ee
where $V_{h}$ is given by \eqref{gravityop}, $V_b$ by \eqref{bvertex}, and $V_{\varphi}$ by \eqref{DilatonVO}. Computing $QV_{\mathrm{NS}}$ is straightforward: we simply add together the results for the BRST operator acting on each of the three components, \eqref{NSgraviton}, \eqref{bfield} and \eqref{dilaton}. The distinct ghost structures in the result impose different constraints on the background fields.

From the terms proportional to $c\tilde{c} \e^{-\phi}\bar{\eta}$ and $c \tilde{c}\e^{-\bar{\phi}}\eta$, we obtain the gauge conditions
\be\label{NSgf}
\nabla^{\mu}h_{\mu\nu}=2\,h_{\mu\nu}\,\partial^{\mu}\Phi\,, \qquad \nabla^{\mu}b_{\mu\nu}=2\,b_{\mu\nu}\,\partial^{\mu}\Phi\,.
\ee
Terms proportional to $c\tilde{c}\partial\tilde{c} \e^{-\phi}\e^{-\bar\phi}$ encode tensorial equations of motion. The symmetric equation, which appears contracted into $(2\bar\psi^{(\mu}\psi^{\nu)}+\partial g^{\mu\nu})$, is
\begin{multline}\label{NSsym}
 \nabla_{\lambda}\nabla^{\lambda}h_{\mu\nu}-2R_{\mu\rho\sigma\nu}\,h^{\rho\sigma}-2R^{\lambda}{}_{(\mu}\,h_{\nu)\lambda}-h^{\rho}_{\sigma}\,H_{\mu\rho\lambda}H_{\nu}{}^{\sigma\lambda} \\
-4\left(h_{\sigma(\mu}\,\nabla_{\nu)}\partial^{\sigma}\Phi+\frac{1}{2}\nabla_{\sigma}h_{\mu\nu}\,\partial^{\sigma}\Phi\right)+H_{\rho\sigma(\mu}\,(\d b)_{\nu)}{}^{\rho\sigma}-4\nabla_{(\mu}\partial_{\nu)}\varphi=0\,,
\end{multline}
while the anti-symmetric equation, which appears contracted into $(\psi^{\mu}\psi^{\nu}-\bar{\psi}^{\mu}\bar\psi^{\nu})$, is
\begin{multline}\label{NSasym}
 \nabla_{\lambda}\nabla^{\lambda}b_{\mu\nu}-2R_{\rho\mu\nu\sigma}\,b^{\rho\sigma}+2R^{\sigma}{}_{[\mu}\, b_{\nu]\sigma} +4\left(b_{\sigma[\mu}\,\nabla_{\nu]}\partial^{\sigma}\Phi-\frac{1}{2}\nabla_{\sigma}b_{\mu\nu}\,\partial^{\sigma}\Phi\right) \\
 +2\nabla_{\rho}h_{\sigma[\nu}\,H_{\mu]}{}^{\rho\sigma}-h^{\rho\sigma}\,\nabla_{\rho}H_{\sigma\mu\nu}-2 H_{\mu\nu\sigma}\,\partial^{\sigma}\varphi=0\,.
\end{multline}
Finally, a scalar equation of motion 
\be\label{NSscalar}
\nabla_{\mu}\partial^{\mu}\varphi-2\partial_{\mu}\Phi\,\partial^{\mu}\varphi - \frac{H\cdot \d b}{24}=0\,
\ee
is imposed by terms proportional to the ghost structure $c\tilde{c}\partial\tilde{c}\,(\e^{-\phi}\partial\e^{-\bar\phi}-\partial\e^{-\phi}\e^{-\bar\phi})$.

Sure enough, equations \eqref{NSgf} are the generalized de Donder gauge conditions for graviton and $B$-field perturbations, while equations \eqref{NSsym} -- \eqref{NSscalar} are precisely the linearised equations of motion for the NS-NS sector of type II supergravity. Thus, $V_{\mathrm{NS}}$ is in the BRST cohomology of the type II ambitwistor string if and only if it encodes a physical, on-shell perturbation for the NS-NS sector of supergravity on space-time.

%%%%%%%%%%%%%%%%%%%%%%%%%%
%%%%%%%%%%%%%%%%%%%%%%%%%%

\section{Discussion}
\label{DiscussionVO}

In this paper, we found vertex operators for the heterotic and type II ambitwistor strings with curved background fields. In the heterotic case, we gave the gluon vertex operator on any Yang-Mills background: BRST closure imposes the physical equations of motion and gauge-fixing constraint on the gluon perturbation. For the type II model things are more subtle. In a pure gravity background, we found graviton and $B$-field vertex operators which are BRST closed when the appropriate physical constraints are imposed on the perturbations. On a general NS-NS background (composed of a metric, $B$-field and dilaton), a fully consistent vertex operator is given by simultaneous encoding perturbations to all three sectors. BRST closure then imposes the appropriate physical constraints on these perturbations, given by the linearised equations of motion and a generalized de Donder gauge.

The fact that these vertex operators can be determined \emph{exactly} -- without recourse to any background field expansion -- points to a significant difference between ambitwistor string and ordinary string theory, where such calculations on a general background would be impossible. It should be noted that a generalization of the vertex operators given here allows for \emph{any} gauge-fixing condition on the perturbations -- the procedure is a straightforward extension of what is done on a flat background~\cite{Berkovits:2018jvm}. The Lorenz or (generalized) de Donder conditions obtained here are, in a sense, the `minimal' such gauge-fixing constraints. 

Of course, one hopes to use these vertex operators to compute physical observables in non-trivial backgrounds. At three-points, this requires knowing the operators in both the fixed (i.e., negative picture number) picture emphasized here, as well as the descended vertex operators (i.e., picture number zero). In the heterotic theory, the descended vertex operator \eqref{Dgluon} is easy to obtain through the standard procedure or linearising the constraint $\mathsf{H}$. 

In the type II case, one can again follow the standard procedure by colliding $V_{\mathrm{NS}}$ with the picture changing operators $\delta(\bar\beta)\cG$ and $\delta(\beta)\bar{\cG}$, respectively. Some terms in the resulting operator will be $Q$-exact and not contribute to correlation functions; these pure gauge contributions can be isolated by applying the picture changing operators in different order, and then comparing the results. Equivalently, the descended vertex operator can be computed by linearising the $\cH$ current \eqref{MostGeneralH} around the chosen background. 

On a general NS-NS background, the resulting vertex operator is complicated, but in highly symmetric backgrounds (usually those of interest for perturbative calculations) the descended vertex operator can be quite tractable. For instance, the three-point graviton amplitude on a vacuum plane wave space-time has been computed directly from ambitwistor strings~\cite{Adamo:2017sze}. We expect the descent procedure to be manageable enough for explicit calculation of 3-point functions around other highly symmetric backgrounds.

To obtain genus zero, $n$-point worldsheet correlations functions (for $n>3$), the analogue of descent with respect to the $\cH$ current must be understood. In flat backgrounds, where $\cH^{\mathrm{flat}}=\Pi^2$, this procedure is understood and leads to the appearance of the scattering equations~\cite{Mason:2013sva,Adamo:2013tsa,Ohmori:2015sha}. However, on general backgrounds $\cH$ has complicated $X$-dependence which obstructs a straightforward evaluation of the path integral. In deformations of the ambitwistor string, where $\cH$ has $X$-dependence even in flat backgrounds, it is still not understood how to perform descent with respect to $\cH$~\cite{Azevedo:2017yjy,Jusinskas:2016qjd,Casali:2016atr}. Clearly, a resolution of this issue is required if ambitwistor strings are to be a useful tool in the study of perturbative QFT on curved backgrounds.

Finally, we note that the fate of the GSO projection (which ensures that the spectrum of the type II ambitwistor string is equivalent to that of type II supergravity) in curved space remains unclear. Indeed, in the graviton vertex operator \eqref{gravityop} the term proportional to a worldsheet derivative does not obey the na\"ive GSO projection, but is clearly required to ensure that $QV_h=0$ yields covariant equations. Other terms in the $B$-field and dilaton vertex operators also na\"ively seem to be in the GSO-odd sector, but dropping them yields non-covariant or unphysical (algebraic and first derivative) equations of motion.

One potential way to address the issue of the GSO projection is to formulate the curved space worldsheet theory with two real fermion systems, rather than the complex fermion system used here. The price to pay is that the action is no longer free and a true background field expansion must be used. OPEs would be calculated order-by-order in perturbation theory, but we expect that calculations of the nilpotency of $Q$ and $Q$-closure of vertex operators will become trivial after a certain low loop order. This follows from the fact that the non-perturbative calculations using the complex fermion model give only a finite number of low order poles in the OPEs.

\acknowledgments

We would like to thank Lionel Mason for useful discussions. TA is supported by an Imperial College Junior Research Fellowship; EC was supported by EPSRC grant EP/ M018911/1; SN is supported by EPSRC grant EP/M50659X/1 and a Studienstiftung des deutschen Volkes scholarship. This research is supported in part by U.S. Department of Energy grant DE-SC0009999 and by funds provided by the University of California.

\bibliography{biblio}
\bibliographystyle{JHEP}

\end{document}